\documentclass[aps,twocolumn,prl,amsmath,amssymb]{revtex4}

\usepackage{graphicx,epsfig}
\usepackage{color}
\usepackage[normalem]{ulem}
\usepackage{bbold}

\begin{document}
\title{Emission Noise in an Interacting Quantum Dot: Role of Inelastic Scattering\\
and Asymmetric Coupling to the Reservoirs}
\author{A.~Cr\'epieux$^{1}$}
\author{S.~Sahoo$^{2,3}$}
\author{T.Q.~Duong$^{1}$}
\author{R.~Zamoum$^{4}$}
\author{M.~Lavagna$^{2,5}$}
\affiliation{$^1$ Aix Marseille Univ, Universit\'e de Toulon, CNRS, CPT UMR 7332, 13288 Marseille, France}
\affiliation{$^2$ Univ. Grenoble Alpes, CEA, INAC-Pheliqs, 38000 Grenoble, France}
\affiliation{$^3$ Physics Department and Research Center OPTIMAS, University of Kaiserslautern, 67663 Kaiserslautern, Germany}
\affiliation{$^4$ Facult\'e des sciences et des sciences appliqu\'ees, Universit\'e de Bouira, rue Drissi Yahia, Bouira 10000, Algeria}
\affiliation{$^5$ Centre National de la Recherche Scientifique -- CNRS, 38042 Grenoble, France} 

\parindent = 0pt

\begin{abstract}
A theory is developed for the emission noise at frequency $\nu$ in a quantum dot in the presence of Coulomb interactions and asymmetric couplings to the reservoirs. We give an analytical expression for the noise in terms of the various transmission amplitudes. Including inelastic scattering contribution, it can be seen as the analog of the Meir-Wingreen formula for the current. A physical interpretation is given on the basis of the transmission of one electron-hole pair to the concerned reservoir where it emits an energy after recombination. We then treat the interactions by solving the self-consistent equations of motion for the Green functions. The results for the noise derivative versus $eV$ show a zero value until $eV = h\nu$, followed by a Kondo peak in the Kondo regime, in good agreement with recent measurements in carbon nanotube quantum dots.
\end{abstract}

\maketitle


In quantum devices, the fluctuations of electrical current provide information on the dynamics of electrons \cite{Landauer1998,Blanter2000,Martin2005,Crepieux2017}, as well as on the energy-photon exchange with the measurement circuit or with the electromagnetic environment \cite{Beenakker,Gabelli2004,Gustavsson2007,Zakka2010,Fulga2010,Lebedev2010,Schneider2010,Schneider2012,Lu2013,Kaasbjerg2015,Forgues2016,Simoneau2017}. Understanding the nature of these fluctuations in a quantum dot (QD) is thus a crucial step insofar as this system is the elementary brick of quantum circuits. The measurement of current fluctuations in a QD is becoming more and more precise, and reliable results are now available both at zero-frequency \cite{Onac2006,Zarchin2008,Ferrier2017} and finite-frequency \cite{Basset2010,Basset2012,Delagrange2017}. Interpreting these experimental findings turned out to be a challenging task, especially in the case of a biased interacting QD with asymmetric couplings to the reservoirs. In view of the challenges, there is an increasing need to develop a theory for calculating the current noise in non-equilibrium, incorporating the inelastic scattering contributions that play a crucial role when Coulomb interactions are present. So far, most of the noise calculations in a QD connected to left ($L$) and right ($R$) reservoirs, either do not distinguish between the noise in the $L$-reservoir and that in the $R$-reservoir \cite{Engel2004,Sela2006,Dong2008,Vitushinsky2008,Mora2008,Mora2009,Moca2011,Muller2013,Moca2014} or assume 
that the left coupling strength $\Gamma_L$ and the right coupling strength $\Gamma_R$ are equal \cite{Sela2006,Hammer2011,Zamoum2016,Zamoum2016bis}, in apparent contradiction with experiments~\cite{Basset2010,Basset2012,Delagrange2017}. Indeed, the measured asymmetry of the couplings can be very large, e.g. $a=11$ \cite{Delagrange2017}, where $a=\Gamma_L/\Gamma_R$ is the asymmetry factor. Certainly, there are theoretical works where the distinction between left and right couplings is made, but these works are limited to the calculations of the zero-frequency noise \cite{Mora2009} and symmetrized noise (generally not the quantity measured in experiments) both for non-interacting \cite{Buttiker1992,Blanter2000,Martin2005,Souza2008,Aguado2004,Marcos2010} and 
interacting QDs~\cite{Ding2013,Droste2015}. In some other works, a linear combination of the auto-correlators and of 
the cross-correlators is calculated~\cite{Rothstein2009,Gabdank2011}. In summary, developing an efficient theory to calculate the finite-frequency noise in non-equilibrium, and investigating the effects of Coulomb interactions and of coupling asymmetry on the noise profile in each reservoir are important unsolved issues which we address in this Letter.


The noise considered here is the emission noise~\cite{Lesovik1997,Aguado2000,Deblock2003} at frequency $\nu$, ${S}_{\alpha\beta}(\nu)=\int_{-\infty}^{\infty} \langle \Delta \hat{I}_\alpha(t) \Delta \hat{I}_\beta(0) \rangle e^{-2i\pi\nu t}dt$, where $\Delta \hat{I}_\alpha(t)=\hat{I}_\alpha(t)-\langle \hat I_\alpha\rangle$ is the deviation of the current from its average value (the index $\alpha$ ($\beta$) represents one of the two reservoirs). We calculate $S_{\alpha\beta}(\nu)$ in an interacting QD by using the non-equilibrium Keldysh Green function technique. When the system is in a steady state, we establish the following formula:
\begin{eqnarray}\label{NS_noise}
\mathcal{S}_{\alpha\beta}(\nu)=\frac{e^2}{h}\sum_{\gamma\delta}\int_{-\infty}^{\infty}d\varepsilon M_{\alpha\beta}^{\gamma\delta}(\varepsilon, \nu)f^e_\gamma(\varepsilon)f^h_\delta(\varepsilon-h\nu)~,
\end{eqnarray}
where $f^{e}_{\gamma}(\varepsilon)$ and $f^{h}_{\delta}(\varepsilon)=1-f^{e}_{\delta}(\varepsilon)$ are the Fermi-Dirac functions for electrons in the $\gamma$-reservoir and holes in the $\delta$-reservoir respectively, and where the matrix elements $M_{\alpha\beta}^{\gamma\delta}(\varepsilon, \nu)$ are listed in Table~I. These elements are written in terms of the transmission amplitude $t_{\alpha\beta}(\varepsilon)$, the transmission coefficient $\mathcal{T}
_{\alpha\beta}(\varepsilon)=|t_{\alpha\beta}(\varepsilon)|^2$, the reflection amplitude $r_{\alpha\alpha}(\varepsilon)=1-t_{\alpha\alpha}(\varepsilon)$, and an effective transmission coefficient defined as $\mathcal{T}_{LR}^{\text{eff},\alpha}(\varepsilon)=2\text{Re}\{t_{\alpha\alpha}(\varepsilon)\}-\mathcal{T}_{\alpha\alpha}(\varepsilon)$~\cite{Note1}. The transmission amplitude is related to the retarded Green function in the QD for spin $\sigma$, $G_\sigma^r(\varepsilon)$, through:  $t_{\alpha\beta}(\varepsilon)=i\sqrt{\Gamma_\alpha\Gamma_\beta}G_\sigma^r(\varepsilon)$, where $\Gamma_\alpha=2\pi\rho_\alpha|V_\alpha|^2$ is the coupling between the QD and the $\alpha$-reservoir,
$V_\alpha$ being the electron hopping amplitude between the QD and the $\alpha$-reservoir, the density of states of which is $\rho_\alpha$. To lighten the notations, we do not put a spin index to $t_{\alpha\beta}(\varepsilon)$ since we consider a spin-unpolarized QD.

\begin{widetext}

\begin{table}[t]
\begin{center}
\begin{tabular}{|c||c|c|c|c|}
\hline
$M_{\alpha\beta}^{\gamma\delta}(\varepsilon,\nu)$& $\gamma=\delta=L$& $\gamma=\delta=R$&$\gamma=L$, $\delta=R$&$\gamma=R$, $\delta=L$\\ \hline\hline
$\alpha=L$&$\mathcal{T}_{LR}^{\text{eff},L}(\varepsilon)\mathcal{T}_{LR}^{\text{eff},L}(\varepsilon-h\nu)$& $\mathcal{T}_{LR}(\varepsilon)\mathcal{T}_{LR}(\varepsilon-h\nu)$ & $[1-\mathcal{T}_{LR}^{\text{eff},L}(\varepsilon)]\mathcal{T}_{LR}(\varepsilon-h\nu)$ & $\mathcal{T}_{LR}(\varepsilon)[1-\mathcal{T}_{LR}^{\text{eff},L}(\varepsilon-h\nu)]$\\
$\beta=L$&$+|t_{LL}(\varepsilon)-t_{LL}(\varepsilon-h\nu)|^2$&&&\\
 \hline
$\alpha=R$& $\mathcal{T}_{LR}(\varepsilon)\mathcal{T}_{LR}(\varepsilon-h\nu)$ &$\mathcal{T}_{LR}^{\text{eff},R}(\varepsilon)\mathcal{T}_{LR}^{\text{eff},R}(\varepsilon-h\nu)$ & $\mathcal{T}_{LR}(\varepsilon)[1-\mathcal{T}_{LR}^{\text{eff},R}(\varepsilon-h\nu)]$ & $[1-\mathcal{T}_{LR}^{\text{eff},R}(\varepsilon)]\mathcal{T}_{LR}(\varepsilon-h\nu)$ \\
$\beta=R$&&$+|t_{RR}(\varepsilon)-t_{RR}(\varepsilon-h\nu)|^2$&&\\
 \hline
$\alpha=L$& $t_{LR}(\varepsilon)t^*_{LR}(\varepsilon - h\nu)$ & $t^*_{LR}(\varepsilon)t_{LR}(\varepsilon - h\nu) $  & $t_{LR}(\varepsilon)t_{LR}(\varepsilon-h\nu)$ &$t_{LR}^*(\varepsilon)t_{LR}^*(\varepsilon-h\nu)$\\
$\beta=R$&$\times[r^*_{LL}(\varepsilon)r_{LL}(\varepsilon-h\nu)-1]$&$\times[r_{RR}(\varepsilon)r^*_{RR}(\varepsilon-h\nu)-1]$&$\times r_{LL}^*(\varepsilon)r_{RR}^*(\varepsilon-h\nu)$& $\times r_{RR}(\varepsilon)r_{LL}(\varepsilon-h\nu)$\\
 \hline
$\alpha=R$& $t^*_{LR}(\varepsilon)t_{LR}(\varepsilon - h\nu)$&$t_{LR}(\varepsilon)t^*_{LR}(\varepsilon - h\nu)$  &$t_{LR}^*(\varepsilon)t_{LR}^*(\varepsilon-h\nu)$&
$t_{LR}(\varepsilon)t_{LR}(\varepsilon-h\nu)$ \\
$\beta=L$&$\times[r_{LL}(\varepsilon)r^*_{LL}(\varepsilon-h\nu)-1]$&$\times[r^*_{RR}(\varepsilon)r_{RR}(\varepsilon-h\nu)-1]$&$\times r_{LL}(\varepsilon)r_{RR}(\varepsilon-h\nu)$& $\times r^*_{RR}(\varepsilon)r^*_{LL}(\varepsilon-h\nu)$\\
 \hline
\end{tabular}
\caption{Expressions of the matrix elements $M_{\alpha\beta}^{\gamma\delta}(\varepsilon, \nu)$ involved in Eq.~(\ref{NS_noise}) for the noise $S_{\alpha\beta}(\nu)$ of an interacting QD with arbitrary coupling symmetry. $\mathcal{T}_{LR}^{\text{eff},\alpha}(\varepsilon)$ is an effective transmission coefficient defined as $\mathcal{T}_{LR}^{\text{eff},\alpha}
(\varepsilon)=2\text{Re}\{t_{\alpha\alpha}(\varepsilon)\}-\mathcal{T}_{\alpha\alpha}(\varepsilon)$.}
\label{table1}
\end{center}
\end{table}

\vspace{-0.5cm}

\begin{figure}[t]
\hspace*{-0.5cm}\includegraphics[width=18cm]{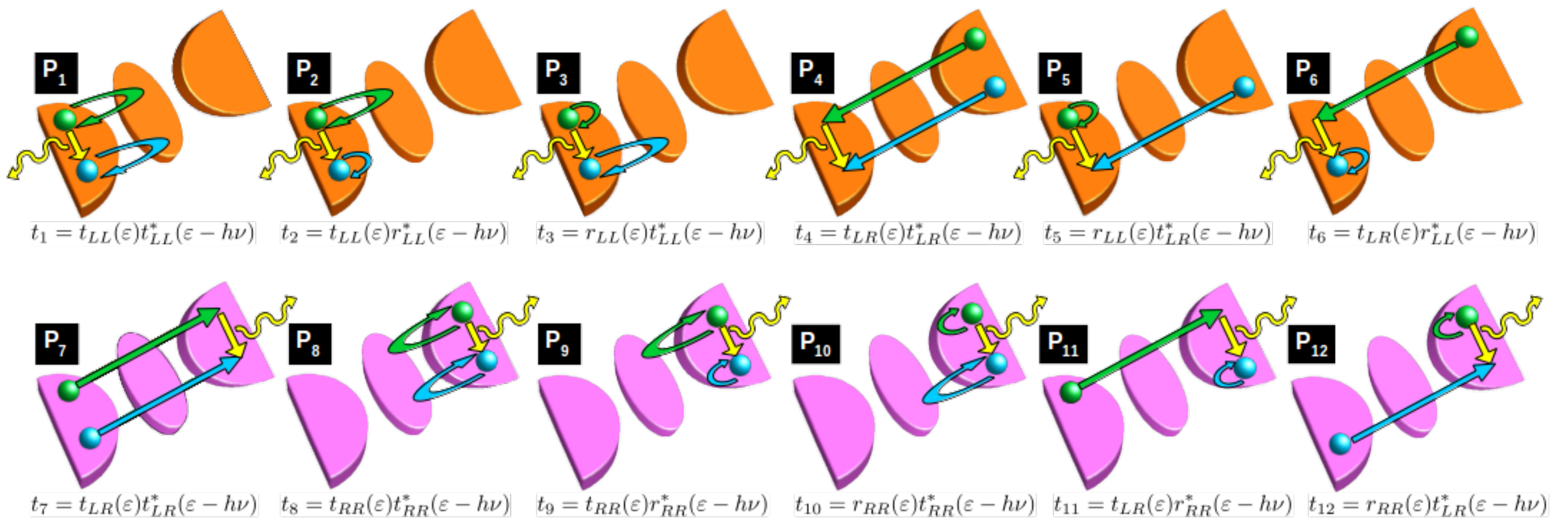}
\caption{Illustration of the six physical processes contributing to $\mathcal{S}_{LL}(\nu)$  with the emission of an energy $h\nu$ in the $L$-reservoir (top row with orange background devices), and the other six physical processes contributing to $\mathcal{S}_{RR}(\nu)$  with the emission of an energy $h\nu$ in the $R$-reservoir (bottom row with pink background devices). The transmission amplitude $t_i$ of the e-h pair for each process with $i\in[1,12]$ is indicated at the bottom of each diagram. A green (blue) sphere represents an electron (a hole) and a yellow wavy arrow represents the emission of energy $h\nu$ in one of the reservoirs.}
\label{figure1}
\end{figure}

\end{widetext}

It is important to underline that even if the noise formula for the interacting QD shows an apparent similarity to the one obtained for the non-interacting QD \cite{Zamoum2016}, the two formulas are distinct since the effective transmission coefficient $\mathcal{T}_{LR}^{\text{eff},\alpha}(\varepsilon)$ differs from $\mathcal{T}_{LR}(\varepsilon)$. Especially $\mathcal{T}_{LR}^{\text{eff},\alpha}(\varepsilon)$ incorporates inelastic scattering contributions~\cite{Zarand2004,Borda2007} which are crucial in interacting systems at
finite temperature and voltage. In the absence of interactions, or when only elastic scattering processes are present, $\mathcal{T}_{LR}^{\text{eff},\alpha}(\varepsilon)$ 
simply equals $\mathcal{T}_{LR}(\varepsilon)$: this result follows from the optical theorem which is verified in this case~\cite{SM}, and allows one to recover the formula established in Ref.~[\onlinecite{Zamoum2016}] in the non-interacting case. The difference between the noise formula in Eq.~(\ref{NS_noise}) and its
non-interacting counterpart can be seen as the exact analog of the difference between the Meir-Wingreen formula
for the current valid in the presence of interactions~\cite{Meir1992} and the Landauer formula obtained using scattering theory.

The proof of Eq.~(\ref{NS_noise}) is the following (see Ref.~[\onlinecite{SM}] for details): we start from Eqs.~(A11)-(A15) of Ref.~[\onlinecite{Zamoum2016}], obtained in the flat wide band limit  for the conduction band after having factorized the two-particle Green function in the QD into a product of single-particle Green functions. Provided that the system is in a steady state, 
we have \cite{Haug2008,Ng1993}: $G_\sigma^{\gtrless}(\varepsilon)=G_\sigma^r(\varepsilon)\Sigma_{\text{tot},\sigma}^{\gtrless}(\varepsilon)G_\sigma^a(\varepsilon)$, where $\Sigma_{\text{tot},\sigma}^{\gtrless}(\varepsilon)$ is the total self-energy \cite{Haug2008,Michelini2017}: $\Sigma_{\text{tot},\sigma}^{\gtrless}(\varepsilon)=\Sigma_{L,\sigma}^{\gtrless}(\varepsilon)+\Sigma_{R,\sigma}^{\gtrless}(\varepsilon)+\Sigma_{\text{int},\sigma}^{\gtrless}(\varepsilon)$,
with $\Sigma_{\alpha,\sigma}^{\gtrless}(\varepsilon)$, the self-energy brought by the coupling with the $\alpha$-reservoir, and $\Sigma_{\text{int},\sigma}^{\gtrless}(\varepsilon)$, 
 the additional self-energy brought by the interactions in the QD. Making use of these relations and noticing that the linear and quadratic terms in $\Sigma_{\text{int},\sigma}^{\gtrless}(\varepsilon)$ cancel in the steady state, one derives Eq.~(\ref{NS_noise}).


In the same way that in the Landauer approach the current is interpreted  in terms of transmission of electrons from $L$-reservoir to $R$-reservoir, the auto-correlator $\mathcal{S}_{\alpha\alpha}(\nu)$ can be interpreted in terms of transmission of e-h pairs or their constituents through the QD, from all possible initial locations, before the pairs recombine leading to the emission of an energy $h\nu$ in the $\alpha$-reservoir. To get $\mathcal{S}_{\alpha\alpha}(\nu)$, we thus have to identify the whole set of such physical processes for each given initial state, determine their transmission amplitudes $t_i$, and take the quantum superposition $|\sum_i t_i|^2$ to calculate the transmission probability. The processes contributing to $\mathcal{S}_{LL}(\nu)$ are six in number as depicted in the top row of~Fig.~\ref{figure1}. We restrict the discussion to $\mathcal{S}_{LL}(\nu)$ because one can straightforwardly deduce $\mathcal{S}_{RR}(\nu)$ by interchanging $L$ and $R$ indices.

When the e-h pair is initially located in the $L$-reservoir, there are three possibilities to emit energy in the $L$-reservoir by recombination of e-h pairs: (i) through process P$_1$ in which one electron of energy $\varepsilon$ (green sphere) and one hole of energy $\varepsilon-h\nu$ (blue sphere) both experience an excursion into the QD and come back to the $L$-reservoir, corresponding to the transmission amplitude $t_1=t_{LL}(\varepsilon)t^*_{LL}(\varepsilon-h\nu)$; (ii) through process P$_2$ in which the electron experiences an excursion into the QD and comes back to the $L$-reservoir, whereas the hole is reflected by the left barrier, corresponding to the transmission amplitude $t_2=t_{LL}(\varepsilon)r^*_{LL}(\varepsilon-h\nu)$; and (iii) through process P$_3$ in which the hole experiences an excursion into the QD and comes back to the $L$-reservoir whereas the electron is reflected, corresponding to the transmission amplitude $t_3=r_{LL}(\varepsilon)t^*_{LL}(\varepsilon-h\nu)$. By taking the quantum superposition of these three processes, $|t_1+t_2+t_3|^2$, we get a contribution to the noise which is equal to the matrix element $M_{LL}^{LL}(\varepsilon,\nu)$ of Table~\ref{table1}~\cite{SM}. Note that even if the amplitudes $t_{1,2,3}$ involve the $L$-index only, we use the subscript $LR$ in the notation for the effective transmission coefficient, $\mathcal{T}_{LR}^{\text{eff},L}(\varepsilon)$, for the reason that it gives back $\mathcal{T}_{LR}(\varepsilon)$ when the optical theorem holds~\cite{SM}.

When the e-h pair is initially located in the $R$-reservoir, both particles cross the entire structure to emit energy in the $L$-reservoir by recombination, as depicted in Fig.~\ref{figure1}(P$_4$), giving rise to the transmission amplitude $t_4=t_{LR}(\varepsilon)t^*_{LR}(\varepsilon-h\nu)$, which leads to the matrix element $M_{LL}^{RR}(\varepsilon,\nu)$ of Table~\ref{table1} after taking $|t_4|^2$. When the electron is initially located in the $L$-reservoir and the hole in the $R$-reservoir, as depicted in Fig.~\ref{figure1}(P$_5$), the electron is reflected and the hole transmitted, giving rise to the transmission amplitude $t_5=r_{LL}(\varepsilon)t^*_{LR}(\varepsilon-h\nu)$ which leads to the matrix element $M_{LL}^{LR}(\varepsilon,\nu)$. By symmetry, the transmission amplitude in process P$_6$ is $t_6=t_{LR}(\varepsilon)r^*_{LL}(\varepsilon-h\nu)$, leading to the matrix element $M_{LL}^{RL}(\varepsilon,\nu)$.  We do not need to take any quantum superposition for the three processes P$_4$-P$_6$ as each of them corresponds to a different initial state.


To get the cross-correlators, one needs to consider the interference terms between the processes accompanied by an emission of energy in both reservoirs \cite{Blanter2000,Martin2005,Hammer2011}. Our study shows that the sum $\mathcal{S}_{LR}(\nu)+\mathcal{S}_{RL}(\nu)$ corresponds to the interference between the processes P$_5$ and P$_{11}$ as regards the term proportional to $f_L^e(\varepsilon)f_R^h(\varepsilon-h\nu)$, since $M_{LR}^{LR}(\varepsilon,\nu)+M_{RL}^{LR}(\varepsilon,\nu)=t_5 t^*_{11}+t_5^* t_{11}$, and to the interference between the processes P$_6$ and P$_{12}$ as regards the term proportional to $f_R^e(\varepsilon)f_L^h(\varepsilon-h\nu)$, since $M_{LR}^{RL}(\varepsilon,\nu)+M_{RL}^{RL}(\varepsilon,\nu)=t_6 t^*_{12}+t_6^* t_{12}$. These interference terms can be either positive or negative according to the relative values of $\varepsilon$ and $\nu$, 
but become strictly negative at zero-frequency due to charge conservation. As far as the contributions proportional to $f_\alpha^e(\varepsilon)f_\alpha^h(\varepsilon-h\nu)$ are concerned, they are given by the interference between the process P$_7$ and the set of processes P$_1$-P$_3$ when $\alpha=L$, and between the process P$_4$ and the set of processes P$_8$-P$_{10}$ when $\alpha=R$.


The noise, given by Eq.~(\ref{NS_noise}) with $M_{\alpha\beta}^{\delta\gamma}(\varepsilon,\nu)$ of Table~I, is completely determined once the retarded Green function $G_\sigma^r(\varepsilon)$ in the QD is known. For the non-interacting single energy level QD, we take the Breit-Wigner form: $G_\sigma^r(\varepsilon)=[\varepsilon-\varepsilon_0+i(\Gamma_L+\Gamma_R)/2]^{-1}$ where $\varepsilon_0$ is the QD energy level. For the interacting single energy level QD, we use the self-consistent renormalized equation-of-motion approach, as developed in Refs.~[\onlinecite{Roermund2010,Lavagna2015,Lavagna2017}], which applies to both equilibrium and non-equilibrium and allows one to determine $G_\sigma^r(\varepsilon)$~\cite{SM}. It has been successfully used \cite{Sahoo2016} to quantitatively explain the experimental results~\cite{Kobayashi210} about the interplay of spin accumulation and magnetic field in a Kondo QD, and is well adapted to describe the Kondo regime in which the noise measurements are performed~\cite{Delagrange2017}. 

\begin{figure}[t]
\centering
\includegraphics[width=8cm]{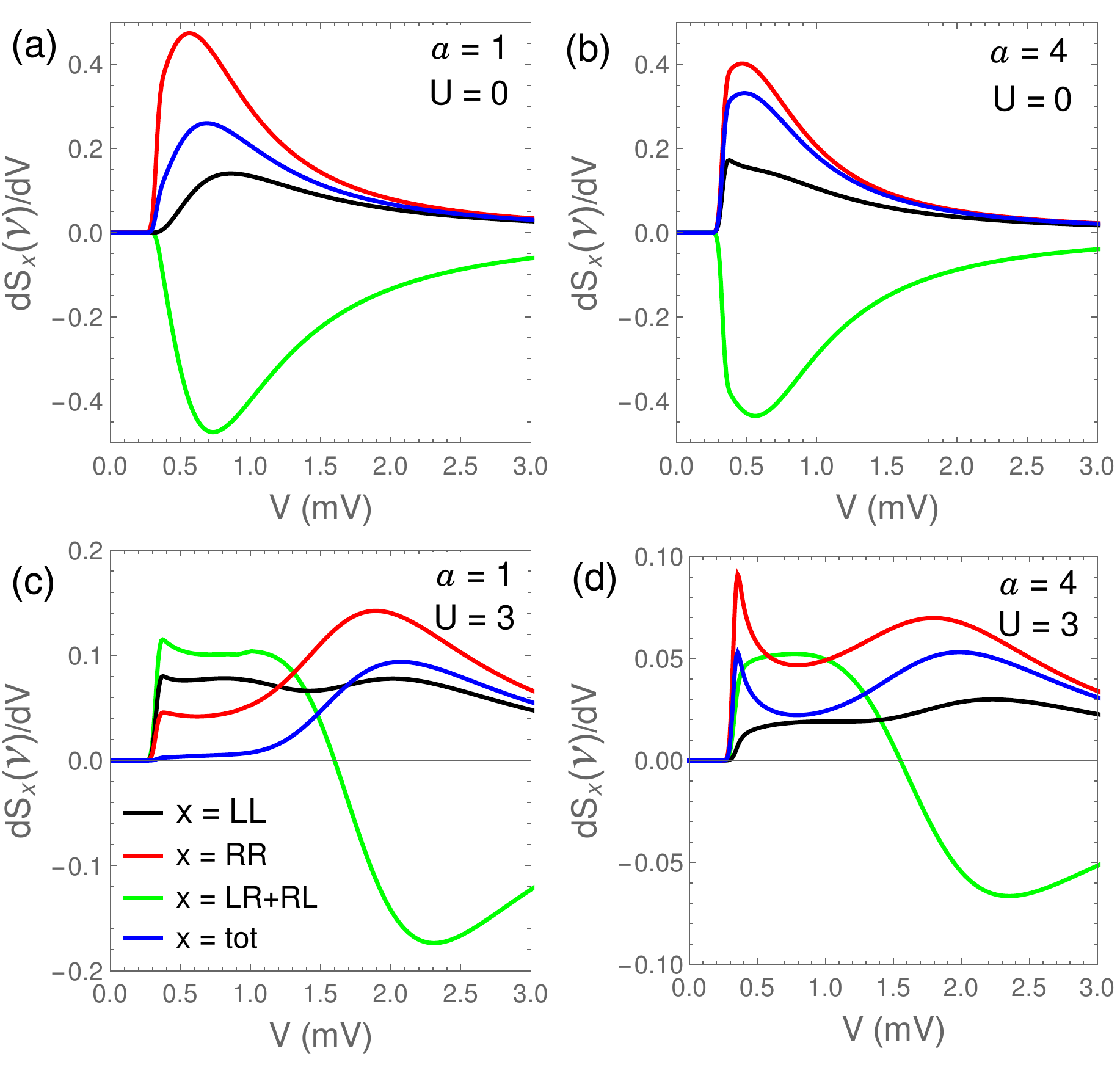}
\caption{Noise derivative $d\mathcal{S}_{\text{x}}(\nu)/dV$ as a function of $V$ (with $\mu_L=0$, $\mu_R=-eV$) at $T=80$ mK, $\nu=78$ GHz (chosen such that $h\nu<k_BT_K$) for $\varepsilon_0=-U/2$ (middle of the Kondo ridge). (a) and (b): $U=0$. (c) and (d): $U=3$ meV. (a) and (c): $\Gamma_{L,R}=0.5$ meV ($a=1$). (b) and (d): $\Gamma_L=0.8$ meV, $\Gamma_R=0.2$ meV ($a=4$). A Kondo peak is observed close to $eV=h\nu$ when $U\ne 0$. Plots for $V<0$ are not shown since $d\mathcal{S}_{\text{x}}(\nu)/dV$ is an odd function in~$V$.}
\label{figure2}
\end{figure}

\begin{figure}[t]
\includegraphics[width=4.cm]{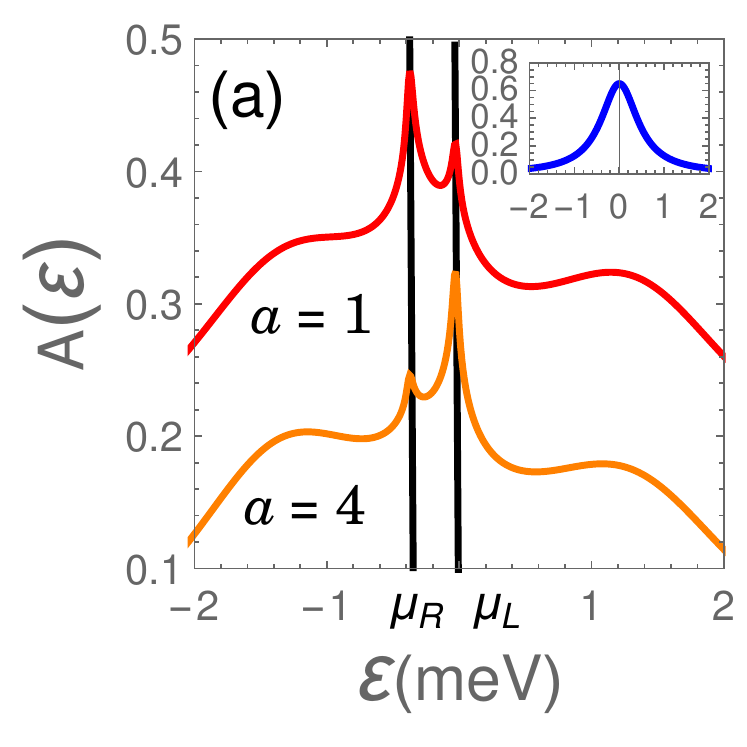}
\hspace{0.4cm}
\includegraphics[width=4.cm]{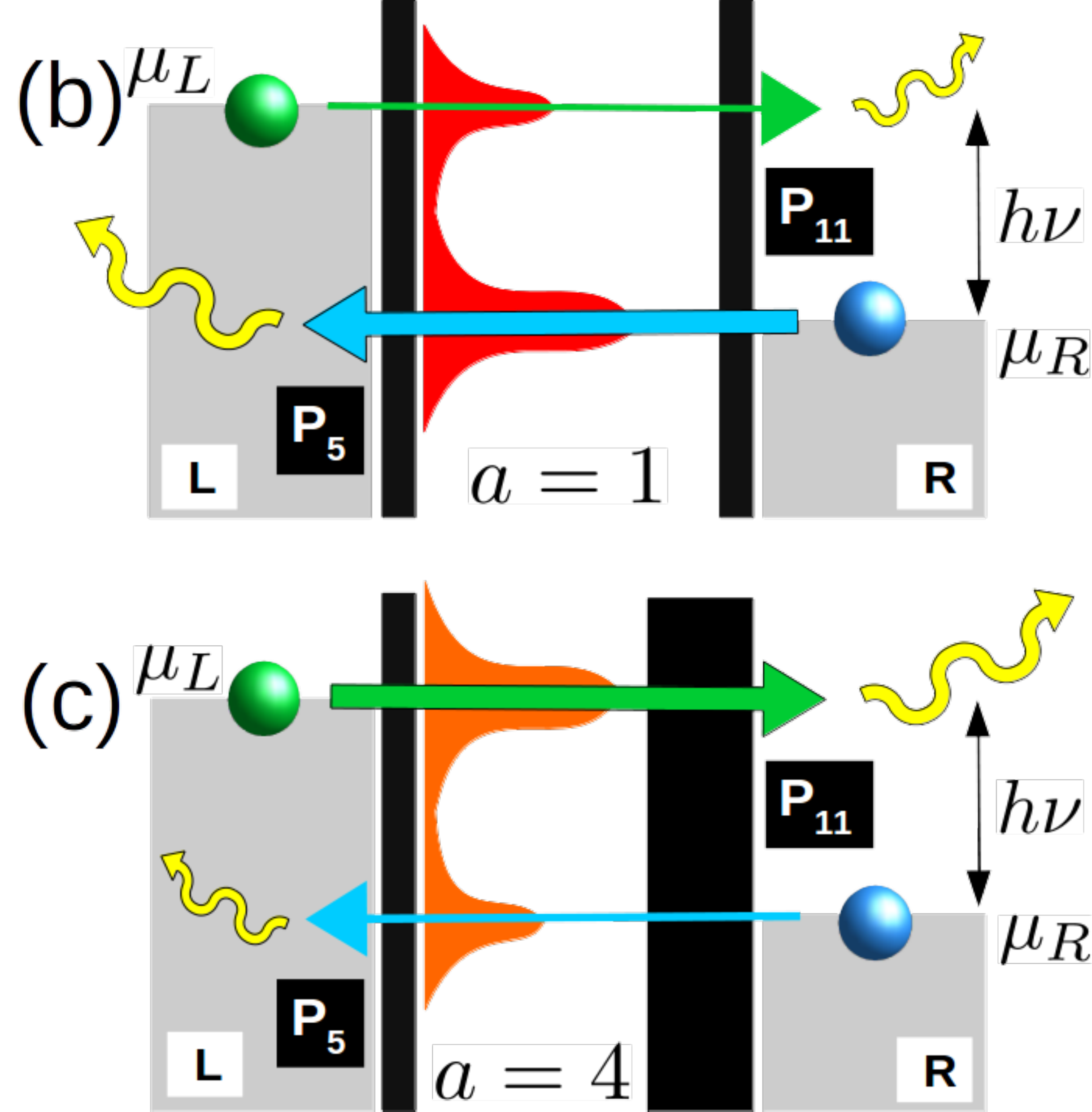}
\caption{{ (a) Spectral density $A(\varepsilon)=-\pi^{-1}\mathrm{Im}\{G_\sigma^r(\varepsilon)\}$ for $\Gamma_{L,R}=0.5$ meV ($a=1$), and $\Gamma_L=0.8$ meV, $\Gamma_R=0.2$ meV ($a=4$), at $U=3$ meV, $\varepsilon_0=-U/2$, and $T=80$ mK. The vertical lines indicate the positions of $\mu_R=-eV=-0.35$ mV and $\mu_L=0$. The $a=1$ curve has been vertically translated for clarity. The spectral density at $U=0$ is shown in blue in inset. (b) and (c) Schematic representation of the  relative importance of the transmission processes at low temperature, low transmission and for $eV\approx h\nu$:  {P$_5$ is dominant over P$_{11}$ for $a=1$, and P$_{11}$ is dominant over~P$_{5}$ for $a=4$.}}}
\label{figure3}
\end{figure}

In Fig.~\ref{figure2}, we report the noise derivative, $d\mathcal{S}_{\alpha\alpha}(\nu)/dV$, as a function of the voltage $V$ for two values of $a=\Gamma_L/\Gamma_R$ and $U$ (with $\varepsilon_0=-U/2$), as well as the derivative of the sum of the cross-correlators $d[\mathcal{S}_\mathrm{LR}(\nu)+\mathcal{S}_\mathrm{RL}(\nu)]/dV$. For completeness, we also plot the derivative of the total noise, $d\mathcal{S}_\mathrm{tot}(\nu)/dV$, where $\mathcal{S}_\mathrm{tot}(\nu)=[\mathcal{S}_{LL}(\nu)+a^2\mathcal{S}_{RR}(\nu)-a\mathcal{S}_{LR}(\nu)-a\mathcal{S}_{RL}(\nu)]/(1+a)^2$, following recent theoretical works which show, by using a current conservation argument along with the Ramo-Shockley theorem, that this is the quantity which is measured in experiments~\cite{Aguado2004,Marcos2010,Droste2015}. A common point to all the curves is the presence of a plateau of value zero at voltage smaller than frequency, here $|V|<h\nu/e = 0.32$ mV, since $h\nu=78$ GHz. The origin of this plateau is related to the fact that the system cannot emit at a frequency higher than the energy provided to it, i.e. the voltage, in full agreement with experiments~\cite{Basset2012,Delagrange2017}. In the absence of interaction (Figs.~\ref{figure2}(a) and (b)), the noise derivatives present a broad peak at $|eV|>h\nu$. Its intensity is larger for $d\mathcal{S}_{RR}(\nu)/dV$ than for $d\mathcal{S}_{LL}(\nu)/dV$ for both symmetric and asymmetric couplings, due to the fact that the $L$-reservoir is grounded ($\mu_L=0$). The effect of the coupling asymmetry is to shift the position of the broad peak towards lower values of $V$. Note that in both cases, the derivative of $\mathcal{S}_{LR}(\nu)+\mathcal{S}_{RL}(\nu)$ is sign negative (green curves in Figs.~\ref{figure2}(a) and (b)). In  the presence of interactions, the electronic transport through a QD is strongly affected. In the Kondo regime, when the number of electrons in the QD is equal to 1 and $T\ll T_K$ ($T_K$ being the Kondo temperature), the differential conductance shows a Kondo peak around $V=0$, in addition to the broad peaks resulting from the Coulomb blockade~\cite{SM}. These effects have their counterparts in the noise. Indeed, the noise derivative shows two clear features (Figs.~\ref{figure2}(c) and (d)): a Kondo peak above $|eV|=h\nu$,  and a secondary broad peak in the proximity of $|eV|=U/2$, corresponding to the boundaries of the Coulomb blockade structure, in full agreement with experiments~\cite{Delagrange2017}. We observe that the noise intensity is reduced in the presence of interactions, as expected in the Kondo regime \cite{Sela2006}. This is related to the fact that  $\mathcal{T}^{\text{eff},\alpha}_{LR}(\varepsilon)>\mathcal{T}_{LR}(\varepsilon)$ (curves not shown), leading to a decrease of $M_{\alpha\alpha}^{LR}(\varepsilon,\nu)$ which provides the dominant contribution at low temperature. Moreover, the derivative of $\mathcal{S}_{LR}(\nu)+\mathcal{S}_{RL}(\nu)$ changes sign at $|eV|=U/2$, going from positive to negative values with increasing $V$. It explains why for symmetric couplings ($a=1$) the total noise derivative becomes smaller below $|eV|=U/2$. We also notice that the height of the Kondo peak in $d\mathcal{S}_{LL}(\nu)/dV$ is larger than in $d\mathcal{S}_{RR}(\nu)/dV$ when $a=1$. This relative order in magnitude is reversed when $\Gamma_L>\Gamma_R$: in this case the Kondo peak becomes more prominent for the more weakly-coupled reservoir. The explanation is the following: 
when $a\ne 1$, (i)~the more pronounced Kondo resonance in the density of states is pinned at the chemical potential ($\mu_L$) of the more strongly-coupled reservoir (orange curve in Fig.~\ref{figure3}(a)), and (ii)~at low temperature, the main process contributing to $d\mathcal{S}_{LL}(\nu)/dV$ is P$_5$ with a probability equal to $\mathcal{T}_{LR}(\varepsilon-h\nu)$ at low transmission, whereas the main process contributing to $d\mathcal{S}_{RR}(\nu)/dV$ is P$_{11}$ with a probability equal to $\mathcal{T}_{LR}(\varepsilon)$ at low transmission. In the process~P$_5$ of Fig.~\ref{figure3}(c), there is a transfer of holes from the $R$-reservoir to the $L$-reservoir at energy close to $\mu_R=-eV$, in the vicinity of which a relatively smaller Kondo resonance is observed, whereas in the process P$_{11}$ of Fig.~\ref{figure3}(c), there is a transfer of electrons from the $L$-reservoir to the $R$-reservoir at energy close to $\mu_L=0$, in the vicinity of which a stronger Kondo resonance is observed (orange curve in Fig.~\ref{figure3}(a)). Since P$_{11}$ contributes to $d\mathcal{S}_{RR}(\nu)/dV$, the Kondo peak is more visible in $d\mathcal{S}_{RR}(\nu)/dV$. In the same way, Fig.~\ref{figure3}(b) illustrates how the height of the Kondo peak in $d\mathcal{S}_{LL}(\nu)/dV$ is larger than that in $d\mathcal{S}_{RR}(\nu)/dV$ when $a=1$.


We have established a general formula for the emission noise in an interacting QD asymmetrically coupled to reservoirs taking the inelastic scattering contributions into account, and we have given a physical interpretation of the results in terms of the transmission of an e-h pair through the QD with an emission of energy. Combining the theory with the equation-of-motion approach to determine the transmission amplitudes entering the noise formula, we have discussed the profile of the noise derivative. The obtained results explain most of the distinctive features recently observed for the noise in a carbon nanotube QD, specially, the presence or the absence of a narrow peak in $d\mathcal{S}_{\text{x}}(\nu)/dV$ versus $V$ in the vicinity of $\pm h \nu/e$, and why the Kondo peak in the noise derivative is more prominent in the more weaky-coupled reservoir. The theory developed in this Letter can be applied to treat other realistic systems.

{\it Acknowledgments} -- We would like to thank H.~Baranger, H.~Bouchiat, R.~Deblock, R.~Delagrange, M.~Guigou, T.~Martin, F.~Michelini and X.~Waintal for valuable discussions. For financial support, the authors acknowledge the Indo-French Centre for the Promotion of Advanced Research (IFCPAR) under Research Project No.4704-02.



\widetext

\pagebreak

\begin{center}
\textbf{Emission Noise in an Interacting Quantum Dot: Role of Inelastic Scattering\\
and Asymmetric Coupling to the Reservoirs -- Supplemental Material}\\~\\
A. Cr\'epieux,$^1$ S. Sahoo,$^{2,3}$ T. Q. Duong,$^1$ R. Zamoum,$^4$ and M. Lavagna$^{2,5}$\\
{\small\it $^1$Aix Marseille Univ, Universit\'e de Toulon, CNRS, CPT UMR 7332, 13288 Marseille, France\\
$^2$Univ. Grenoble Alpes, CEA, INAC-Pheliqs, 38000 Grenoble, France\\
$^3$Physics Department and Research Center OPTIMAS, University of Kaiserslautern, 67663 Kaiserslautern, Germany\\
$^4$Facult\'e des sciences et des sciences appliqu\'ees, Universit\'e de Bouira, rue Drissi Yahia, Bouira 10000, Algeria\\
$^5$Centre National de la Recherche Scientifique -- CNRS, 38042 Grenoble, France\\}
\end{center}

\setcounter{equation}{0}
\setcounter{figure}{0}
\setcounter{table}{1}
\setcounter{page}{1}
\makeatletter
\renewcommand{\theequation}{S\arabic{equation}}
\renewcommand{\thefigure}{S\arabic{figure}}

In this Supplemental Material (SM), we first present the detailed calculation of the finite-frequency noise for an interacting quantum dot (QD) with asymmetric couplings to the reservoirs (Section A), secondly we prove the relation $M_{LL}^{LL}(\varepsilon,\nu)=|t_1+t_2+t_3|^2$ (Section B), where the second formulation appears when we take the coherent superposition of the processes P$_1$, P$_2$ and P$_3$. Thirdly, we give the relation between the various transmission amplitudes and coefficients in the case of a non-interacting QD, or when only elastic scattering is present, for which the optical theorem holds, and we derive the noise matrix elements in that case (Section C). Fourthly we give the expression of the self-consistent equations of motion used to determine numerically the retarded Green function in the case of an interacting QD (Section D), and end up by discussing the results obtained for the differential conductance in the case of symmetric and asymmetric couplings (Section E).


\section{A -- Calculation of the current noise for an interacting QD with arbitrary coupling symmetry}

We start from the expression for the current noise in a QD given by Eqs.~(A11-A15) in Ref.~[\onlinecite{Zamoum2016}] obtained in the flat wide band limit (FWBL) for the conduction band. We get $\mathcal{S}_{\alpha\beta}(\nu)=\sum_{i=1}^5\mathcal{C}_{\alpha\beta}^{(i)}(\nu)$ with
\begin{eqnarray}\label{noise1}
\mathcal{C}_{\alpha\beta}^{(1)}(\nu)&=&\frac{e^2}{h}\delta_{\alpha\beta}\int_{-\infty}^{\infty} d\varepsilon \Big[
G_\sigma^<(\varepsilon)\Sigma^>_{\alpha,\sigma}(\varepsilon-h\nu)+\Sigma^<_{\alpha,\sigma}(\varepsilon)G_\sigma^>(\varepsilon-h\nu)\Big]~, \\
\mathcal{C}_{\alpha\beta}^{(2)}(\nu)&=&-\frac{e^2}{h}\int_{-\infty}^{\infty} d\varepsilon \Big[G_\sigma^r(\varepsilon)\Sigma^<_{\alpha,\sigma}(\varepsilon)+G_\sigma^<(\varepsilon)\Sigma^a_{\alpha,\sigma}(\varepsilon)\Big]\Big[G_\sigma^r(\varepsilon-h\nu)\Sigma^>_{\beta,\sigma}(\varepsilon-h\nu)+G_\sigma^>(\varepsilon-h\nu)\Sigma^a_{\beta,\sigma}(\varepsilon-h\nu)\Big]~,\\
\mathcal{C}_{\alpha\beta}^{(3)}(\nu)&=&\frac{e^2}{h}\int_{-\infty}^{\infty} d\varepsilon\Big[\Sigma^<_{\alpha,\sigma}(\varepsilon)G^r_{\sigma}(\varepsilon)\Sigma^r_{\beta,\sigma}(\varepsilon)+\Sigma^a_{\alpha,\sigma}(\varepsilon)G^<_{\sigma}(\varepsilon)\Sigma^r_{\beta,\sigma}(\varepsilon)+\Sigma^a_{\alpha,\sigma}(\varepsilon)G^a_{\sigma}(\varepsilon)\Sigma^<_{\beta,\sigma}(\varepsilon)\Big]G_\sigma^>(\varepsilon-h\nu)~,\\
\mathcal{C}_{\alpha\beta}^{(4)}(\nu)&=&\frac{e^2}{h}\int_{-\infty}^{\infty} d\varepsilon G_\sigma^<(\varepsilon)\Big[\Sigma^r_{\alpha,\sigma}(\varepsilon-h\nu)G^r_{\sigma}(\varepsilon-h\nu)\Sigma^>_{\beta,\sigma}(\varepsilon-h\nu)\nonumber\\
&&+\Sigma^r_{\alpha,\sigma}(\varepsilon-h\nu)G^>_{\sigma}(\varepsilon-h\nu)\Sigma^a_{\beta,\sigma}(\varepsilon-h\nu)+\Sigma^>_{\alpha,\sigma}(\varepsilon-h\nu)G^a_{\sigma}(\varepsilon-h\nu)\Sigma^a_{\beta,\sigma}(\varepsilon-h\nu)\Big]~,\\\label{noise5}
\mathcal{C}_{\alpha\beta}^{(5)}(\nu)&=&-\frac{e^2}{h}\int_{-\infty}^{\infty} d\varepsilon \Big[G_\sigma^<(\varepsilon)\Sigma^r_{\beta,\sigma}(\varepsilon)+G_\sigma^a(\varepsilon)\Sigma^<_{\beta,\sigma}(\varepsilon)\Big]\Big[G_\sigma^>(\varepsilon-h\nu)\Sigma^r_{\alpha,\sigma}(\varepsilon-h\nu)+G_\sigma^a(\varepsilon-h\nu)\Sigma^>_{\alpha,\sigma}(\varepsilon-h\nu)\Big] \bigg]\nonumber~,\\
\end{eqnarray}
where $\Sigma_{\alpha,\sigma}^{r,a,\gtrless}(\varepsilon)=\sum_{k\in\alpha}|V_{\alpha} |^2 g_{k,\alpha,\sigma}^{r,a,\gtrless}(\varepsilon)$ is the contribution to the self-energy brought by the tunneling between the $\alpha$-reservoir and the QD.  $G_\sigma^{r,a,\gtrless}(\varepsilon)$ and $g_{k,\alpha,\sigma}^{r,a,\gtrless}(\varepsilon)$ are the retarded, advanced, Keldysh greater and lesser Green functions in the QD and in the disconnected $\alpha$-reservoir respectively. The approximation made to get Eqs.~(\ref{noise1}-\ref{noise5}) amounts to having factorized the two-particle Green functions in the QD into a product of two single-particle Green functions. Provided that this approximation is made together with the FWBL assumption, the latter expression for the noise is general and valid for any value of the Coulomb interactions in the QD and any symmetry of the tunneling couplings of the dot to the two reservoirs. We remark that even if the spin index $\sigma$ appears in the Green function and self-energy notations, it does not appear as an index in the noise notation since the system we consider in this Letter is spin unpolarized. When the system is in a steady state, $G^{\gtrless}_\sigma(\varepsilon)$ is simply given by \cite{Ng1993,Haug2008} 
\begin{eqnarray}\label{Glesser}
G_\sigma^{\gtrless}(\varepsilon)=G_\sigma^r(\varepsilon)\Sigma_{\text{tot},\sigma}^{\gtrless}(\varepsilon)G_\sigma^a(\varepsilon)~.
\end{eqnarray}

In the presence of interactions, the total self-energy can be put in the form \cite{Haug2008,Michelini2017}
\begin{eqnarray}\label{Sigmatot}
\Sigma_{\text{tot},\sigma}^{r,a,\gtrless}(\varepsilon)=\Sigma_{L,\sigma}^{r,a,\gtrless}(\varepsilon)+\Sigma_{R,\sigma}^{r,a,\gtrless}(\varepsilon)+\Sigma_{\text{int},\sigma}^{r,a,\gtrless}(\varepsilon)~,
\end{eqnarray}
where $\Sigma_{\text{int},\sigma}^{r,a,\gtrless}
(\varepsilon)$ is the additional contribution brought by the interactions residing in the central region. In the 
FWBL, $\Sigma^{r}_{\alpha,\sigma}(\varepsilon)=-i\Gamma_{\alpha} /2$, $\Sigma^{a}_{\alpha,\sigma}(\varepsilon)=i\Gamma_{\alpha} /2$, $\Sigma_{\alpha,\sigma}^<(\varepsilon)=i\Gamma_{\alpha} f_{\alpha}^e(\varepsilon)$, and $\Sigma_{\alpha,\sigma}
^>(\varepsilon)=-i\Gamma_{\alpha}f_{\alpha}^h(\varepsilon)$, where $\Gamma_\alpha=2\pi\rho_\alpha|V_\alpha |^2$, $\rho_\alpha$ being the density of states of the $\alpha$-reservoir and $V_\alpha $ the electron hopping amplitude between the QD and the $\alpha$-reservoir.

In the absence of interactions, $\Sigma_{\text{int},\sigma}(\varepsilon)=0$ and the self-energy contains only the tunneling contributions. As a result, the following relation holds: $G_\sigma^r(\varepsilon)-G_\sigma^a(\varepsilon)=-iG^r_\sigma(\varepsilon)(\Gamma_L+\Gamma_R)G_\sigma^a(\varepsilon)$, ensuring the optical theorem to be satisfied; the expression for the noise in this case is given in the Section C of this SM.

In the general case, when interactions are present, the latter 
relation no longer holds due to the contribution of the interaction self-energy. However the calculation of the noise can still be done according to the procedure presented below provided that the system is in a steady state. By incorporating Eqs.~(\ref{Glesser}) and (\ref{Sigmatot}) into the expression of the noise given in Eqs.~(\ref{noise1}-\ref{noise5}), one gets  
\begin{eqnarray}\label{noise_SM}
\mathcal{S}_{\alpha\beta}(\nu)&=&\frac{e^2}{h}\Gamma_\alpha\delta_{\alpha\beta}\int_{-\infty}^{\infty} d\varepsilon \bigg[
f^h_\alpha(\varepsilon-h\nu)G_\sigma^r(\varepsilon)\sum_\gamma\Gamma_\gamma f_\gamma^e(\varepsilon)G_\sigma^a(\varepsilon)
+f^e_\alpha(\varepsilon)G_\sigma^r(\varepsilon-h\nu)\sum_\gamma\Gamma_\gamma f_\gamma^h(\varepsilon-h\nu)G_\sigma^a(\varepsilon-h\nu)\bigg] \nonumber \\
&&+\frac{e^2}{h}\Gamma_\alpha \Gamma_\beta\int_{-\infty}^{\infty} d\varepsilon \bigg[G_\sigma^r(\varepsilon)\sum_\gamma\Gamma_\gamma f_\gamma^e(\varepsilon)G_\sigma^a(\varepsilon)G_\sigma^r(\varepsilon-h\nu)\sum_\delta\Gamma_\delta f_\delta^h(\varepsilon-h\nu)G_\sigma^a(\varepsilon-h\nu)\nonumber \\
&&-f^e_\alpha(\varepsilon)f^h_\beta(\varepsilon-h\nu) G_\sigma^r(\varepsilon)G_\sigma^r(\varepsilon-h\nu)
-f^e_\beta(\varepsilon)f^h_\alpha(\varepsilon-h\nu)G_\sigma^a(\varepsilon) G_\sigma^a(\varepsilon-h\nu) \nonumber \\
&&-i \big[f^e_\alpha(\varepsilon)G_\sigma^r(\varepsilon)-f^e_\beta(\varepsilon)G_\sigma^a(\varepsilon)\big]G_\sigma^r(\varepsilon-h\nu)\sum_\gamma\Gamma_\gamma f_\gamma^h(\varepsilon-h\nu)G_\sigma^a(\varepsilon-h\nu)\nonumber \\
&&+i\big[f^h_\alpha(\varepsilon-h\nu)G_\sigma^a(\varepsilon-h\nu)
-f^h_\beta(\varepsilon-h\nu) G_\sigma^r(\varepsilon-h\nu)\big]G_\sigma^r(\varepsilon)\sum_\gamma\Gamma_\gamma f_\gamma^e(\varepsilon)G_\sigma^a(\varepsilon)\bigg]~.
\end{eqnarray}
The r.h.s. of Eq.~(\ref{noise_SM}) results from the contribution proportional either to $\Sigma_{\gamma,\sigma}^{<}(\varepsilon) \Sigma_{\delta,\sigma}^{>}(\varepsilon-h\nu)$ or to $\Sigma_{\gamma,\sigma}^{>}(\varepsilon) \Sigma_{\delta,\sigma}^{<}(\varepsilon-h\nu)$ in Eqs.~(\ref{noise1}-\ref{noise5}), once we have inserted Eqs.~(\ref{Glesser}) and (\ref{Sigmatot}). We have 
checked that the remaining contributions coming from the linear and quadratic terms in $\Sigma_{\text{int},\sigma}^{\gtrless}(\varepsilon)$, $\Sigma_{\text{int},\sigma}^{\gtrless}(\varepsilon-h\nu)$ cancel in the steady state.\\

\subsection{Expression for the auto-correlators $\mathcal{S}_{LL}(\nu)$ and $\mathcal{S}_{RR}(\nu)$}

From Eq.~(\ref{noise_SM}), taking $\alpha=\beta=L$ and rearranging the various terms, we get the expression of the auto-correlator noise associated with the $L$-reservoir, that is
\begin{eqnarray}
\mathcal{S}_{LL}(\nu)&=&\frac{e^2}{h}\int_{-\infty}^{\infty} d\varepsilon  \bigg[
\sum_\gamma\Gamma_L\Gamma_\gamma G_\sigma^r(\varepsilon) G_\sigma^a(\varepsilon)f_\gamma^e(\varepsilon)f^h_L(\varepsilon-h\nu) \nonumber \\
&&+\sum_\gamma\Gamma_L\Gamma_\gamma G_\sigma^r(\varepsilon-h\nu) G_\sigma^a(\varepsilon-h\nu)f^e_L(\varepsilon)f_\gamma^h(\varepsilon-h\nu) \nonumber \\
&&+\sum_{\gamma,\delta}\Gamma_L^2\Gamma_\gamma\Gamma_\delta G_\sigma^r(\varepsilon)G_\sigma^a(\varepsilon)G_\sigma^r(\varepsilon-h\nu)G_\sigma^a(\varepsilon-h\nu) f_\gamma^e(\varepsilon)f_\delta^h(\varepsilon-h\nu)\nonumber \\
&&-\Gamma_L^2\big[G_\sigma^r(\varepsilon)G_\sigma^r(\varepsilon-h\nu)+G_\sigma^a(\varepsilon) G_\sigma^a(\varepsilon-h\nu) \big]f^e_L(\varepsilon)f^h_L(\varepsilon-h\nu)  \nonumber \\
&&-i\sum_\gamma\Gamma_L^2 \Gamma_\gamma\big[G_\sigma^r(\varepsilon)-G_\sigma^a(\varepsilon)\big]G_\sigma^r(\varepsilon-h\nu) G_\sigma^a(\varepsilon-h\nu)f^e_L(\varepsilon)f_\gamma^h(\varepsilon-h\nu)\nonumber \\
&&+i\sum_\gamma\Gamma_L^2\Gamma_\gamma\big[G_\sigma^a(\varepsilon-h\nu)
- G_\sigma^r(\varepsilon-h\nu)\big]G_\sigma^r(\varepsilon)G_\sigma^a(\varepsilon) f_\gamma^e(\varepsilon)f^h_L(\varepsilon-h\nu) \bigg]~.
\end{eqnarray}
Introducing the transmission amplitude $t_{\alpha\beta}(\varepsilon)=i\sqrt{\Gamma_\alpha\Gamma_\beta}G_\sigma^r(\varepsilon)$ and the transmission coefficient $\mathcal{T}_{\alpha\beta}(\varepsilon)=|t_{\alpha\beta}(\varepsilon)|^2=\Gamma_\alpha
\Gamma_\beta G_\sigma^r(\varepsilon)G_\sigma^a(\varepsilon)$ in this expression, and performing the sum over $\gamma$ and $\delta$, we get
\begin{eqnarray}
\mathcal{S}_{LL}(\nu)&=&\frac{e^2}{h}\int_{-\infty}^{\infty} d\varepsilon \bigg[
\mathcal{T}_{LR}(\varepsilon)\mathcal{T}_{LR}(\varepsilon-h\nu)f^e_R(\varepsilon)f^h_R(\varepsilon-h\nu) \nonumber \\
&&
+\Big[\mathcal{T}_{LL}(\varepsilon)+\mathcal{T}_{LL}(\varepsilon-h\nu)+\mathcal{T}_{LL}(\varepsilon)\mathcal{T}_{LL}(\varepsilon-h\nu)+t_{LL}(\varepsilon)t_{LL}(\varepsilon-h\nu)+t^*_{LL}(\varepsilon)t^*_{LL}(\varepsilon-h\nu)\nonumber\\
&&-2\text{Re}\{t_{LL}(\varepsilon)\}\mathcal{T}_{LL}(\varepsilon-h\nu)-2\text{Re}\{t_{LL}(\varepsilon-h\nu)\}\mathcal{T}_{LL}(\varepsilon)\Big]f^e_L(\varepsilon)f^h_L(\varepsilon-h\nu) \nonumber \\
&&
+\Big[1-\big[2\text{Re}\{t_{LL}(\varepsilon)\}-\mathcal{T}_{LL}(\varepsilon)\big]\Big]\mathcal{T}_{LR}(\varepsilon-h\nu)f^e_L(\varepsilon)f^h_R(\varepsilon-h\nu) \nonumber \\
&&
+\Big[1-\big[2\text{Re}\{t_{LL}(\varepsilon-h\nu)\}-\mathcal{T}_{LL}(\varepsilon-h\nu)\big]\Big]\mathcal{T}_{LR}(\varepsilon)f^e_R(\varepsilon)f^h_L(\varepsilon-h\nu) \bigg]~.
\end{eqnarray}
We have used this result to write the expression of the matrix elements $M_{LL}^{\gamma\delta}(\varepsilon,\nu)$ along the first row of Table~I introducing an effective transmission coefficient defined as: $\mathcal{T}_{LR}^{\text{eff},L}(\varepsilon)=2\text{Re}\{t_{LL}(\varepsilon)\}-\mathcal{T}_{LL}(\varepsilon)$. The expression for the auto-correlator noise in the $R$-reservoir, $\mathcal{S}_{RR}(\nu)$ given along the second row of Table~I, is obtained from $\mathcal{S}_{LL}(\nu)$ by interchanging the indices $L$ and $R$.

\subsection{Expression for the cross-correlators $\mathcal{S}_{LR}(\nu)$ and $\mathcal{S}_{RL}(\nu)$}

From Eq.~(\ref{noise_SM}), taking $\alpha=L$ and $\beta=R$ and rearranging the various terms, we get the expression for the cross-correlator noise, that is
\begin{eqnarray}
\mathcal{S}_{LR}(\nu)&=&
\frac{e^2}{h}\Gamma_L\Gamma_R\int_{-\infty}^{\infty} d\varepsilon \bigg[\sum_{\gamma,\delta}\Gamma_\gamma \Gamma_\delta G_\sigma^r(\varepsilon)G_\sigma^a(\varepsilon)G_\sigma^r(\varepsilon-h\nu) G_\sigma^a(\varepsilon-h\nu)f_\gamma^e(\varepsilon)f_\delta^h(\varepsilon-h\nu)\nonumber \\
&&-G_\sigma^r(\varepsilon)G_\sigma^r(\varepsilon-h\nu)f^e_L(\varepsilon)f^h_R(\varepsilon-h\nu) 
-G_\sigma^a(\varepsilon) G_\sigma^a(\varepsilon-h\nu)f^e_R(\varepsilon)f^h_L(\varepsilon-h\nu) \nonumber \\
&&-i \sum_\gamma\Gamma_\gamma\big[f^e_L(\varepsilon)G_\sigma^r(\varepsilon)-f^e_R(\varepsilon)G_\sigma^a(\varepsilon)\big]G_\sigma^r(\varepsilon-h\nu)G_\sigma^a(\varepsilon-h\nu) f_\gamma^h(\varepsilon-h\nu)\nonumber \\
&&+i\sum_\gamma\Gamma_\gamma\big[f^h_L(\varepsilon-h\nu)G_\sigma^a(\varepsilon-h\nu)
-f^h_R(\varepsilon-h\nu) G_\sigma^r(\varepsilon-h\nu)\big]G_\sigma^r(\varepsilon) G_\sigma^a(\varepsilon)f_\gamma^e(\varepsilon)\bigg]~.
\end{eqnarray}
Introducing the transmission amplitude and coefficients, and performing the sum over $\gamma$ and $\delta$, we get
\begin{eqnarray}
\mathcal{S}_{LR}(\nu)&=&
\frac{e^2}{h}\int_{-\infty}^{\infty} d\varepsilon \bigg[\Big[\mathcal{T}_{LR}(\varepsilon)\mathcal{T}_{LL}(\varepsilon-h\nu)-t_{LL}(\varepsilon)\mathcal{T}_{LR}(\varepsilon-h\nu)-\mathcal{T}_{LR}(\varepsilon)t^*_{LL}(\varepsilon-h\nu)\Big]f_L^e(\varepsilon)f_L^h(\varepsilon-h\nu)\nonumber \\
&&+\Big[\mathcal{T}_{LR}(\varepsilon)\mathcal{T}_{RR}(\varepsilon-h\nu)-t^*_{RR}(\varepsilon)\mathcal{T}_{LR}(\varepsilon-h\nu)-\mathcal{T}_{LR}(\varepsilon)t_{RR}(\varepsilon-h\nu)\Big]f_R^e(\varepsilon)f_R^h(\varepsilon-h\nu) \nonumber \\
&&+\Big[\mathcal{T}_{LR}(\varepsilon)\mathcal{T}_{LR}(\varepsilon-h\nu)+t_{LR}(\varepsilon)t_{LR}(\varepsilon-h\nu)-t_{RR}(\varepsilon)\mathcal{T}_{LR}(\varepsilon-h\nu)-\mathcal{T}_{LR}(\varepsilon)t_{LL}(\varepsilon-h\nu)\Big]f_L^e(\varepsilon)f_R^h(\varepsilon-h\nu)\nonumber \\
&&+\Big[\mathcal{T}_{LR}(\varepsilon)\mathcal{T}_{LR}(\varepsilon-h\nu)+t^*_{LR}(\varepsilon)t^*_{LR}(\varepsilon-h\nu)-t^*_{LL}(\varepsilon)\mathcal{T}_{LR}(\varepsilon-h\nu)-\mathcal{T}_{LR}(\varepsilon)t^*_{RR}(\varepsilon-h\nu)\Big]f_R^e(\varepsilon)f_L^h(\varepsilon-h\nu)\bigg]~.\nonumber\\
\end{eqnarray}

We have used this result to write the expression of the matrix elements $M_{LR}^{\gamma\delta}(\varepsilon,\nu)$ along the third row of Table~I. The cross-correlator $\mathcal{S}_{RL}(\nu)$, given along the fourth row of Table~I, is obtained from the expression of $\mathcal{S}_{LR}(\nu)$ by interchanging the indices $L$ and $R$.


\section{B -- Proof of the relation: $M_{LL}^{LL}(\varepsilon,\nu)=|t_1+t_2+t_3|^2$}

From Table~I, we have
\begin{eqnarray}
M_{LL}^{LL}(\varepsilon,\nu)&=&|t_{LL}(\varepsilon)-t_{LL}(\varepsilon-h\nu)|^2+\mathcal{T}_{LR}^{\text{eff},L}(\varepsilon)\mathcal{T}_{LR}^{\text{eff},L}(\varepsilon-h\nu)~,
\end{eqnarray}
with $\mathcal{T}_{LR}^{\text{eff},L}(\varepsilon)=2\text{Re}\{t_{LL}(\varepsilon)\}-\mathcal{T}_{LL}(\varepsilon)$, thus
\begin{eqnarray}
M_{LL}^{LL}(\varepsilon,\nu)
&=&|t_{LL}(\varepsilon)-t_{LL}(\varepsilon-h\nu)|^2+\big[2\text{Re}\{t_{LL}(\varepsilon)\}-\mathcal{T}_{LL}(\varepsilon)\big]\big[2\text{Re}\{t_{LL}(\varepsilon-h\nu)\}-\mathcal{T}_{LL}(\varepsilon-h\nu)\big]\nonumber\\
&=&|t_{LL}(\varepsilon)-t_{LL}(\varepsilon-h\nu)|^2+\big[t_{LL}(\varepsilon)+t^*_{LL}(\varepsilon)\big]\big[t_{LL}(\varepsilon-h\nu)+t^*_{LL}(\varepsilon-h\nu)\big]\nonumber\\
&&+\mathcal{T}_{LL}(\varepsilon)\mathcal{T}_{LL}(\varepsilon-h\nu)-2\text{Re}\{t_{LL}(\varepsilon)\}\mathcal{T}_{LL}(\varepsilon-h\nu)-2\text{Re}\{t_{LL}(\varepsilon-h\nu)\}\mathcal{T}_{LL}(\varepsilon)~.
\end{eqnarray}
Knowing that $|t_{LL}(\varepsilon)-t_{LL}(\varepsilon-h\nu)|^2=\mathcal{T}_{LL}(\varepsilon)+\mathcal{T}_{LL}(\varepsilon-h\nu)-t_{LL}^*(\varepsilon)t_{LL}(\varepsilon-h\nu)-t_{LL}(\varepsilon)t_{LL}^*(\varepsilon-h\nu)$, this leads to
\begin{eqnarray}
M_{LL}^{LL}(\varepsilon,\nu)
&=&\mathcal{T}_{LL}(\varepsilon)+\mathcal{T}_{LL}(\varepsilon-h\nu)+t_{LL}(\varepsilon)t_{LL}(\varepsilon-h\nu)
+t_{LL}^*(\varepsilon)t_{LL}^*(\varepsilon-h\nu)\nonumber\\
&&+\mathcal{T}_{LL}(\varepsilon)\mathcal{T}_{LL}(\varepsilon-h\nu)-2\text{Re}\{t_{LL}(\varepsilon)\}\mathcal{T}_{LL}(\varepsilon-h\nu)-2\text{Re}\{t_{LL}(\varepsilon-h\nu)\}\mathcal{T}_{LL}(\varepsilon)~,
\end{eqnarray}
which can be factorized under the form
\begin{eqnarray}\label{MLL}
M_{LL}^{LL}(\varepsilon,\nu)
&=&\mathcal{T}_{LL}(\varepsilon)+[1-t_{LL}(\varepsilon)][1-t_{LL}^*(\varepsilon)]\mathcal{T}_{LL}(\varepsilon-h\nu)\nonumber\\
&&+t_{LL}(\varepsilon)[1-t_{LL}^*(\varepsilon)]t_{LL}(\varepsilon-h\nu)+t^*_{LL}(\varepsilon)[1-t_{LL}(\varepsilon)]t^*_{LL}(\varepsilon-h\nu)\nonumber\\
&=&\mathcal{T}_{LL}(\varepsilon)+\mathcal{R}_{LL}(\varepsilon)\mathcal{T}_{LL}(\varepsilon-h\nu)+t_{LL}(\varepsilon)r^*_{LL}(\varepsilon)t_{LL}(\varepsilon-h\nu)+t^*_{LL}(\varepsilon)r_{LL}(\varepsilon)t^*_{LL}(\varepsilon-h\nu)~,
\end{eqnarray}
where we have used the definitions of the reflection amplitude: $r_{LL}(\varepsilon)=1-t_{LL}(\varepsilon)$, and the reflection coefficient: $\mathcal{R}_{LL}(\varepsilon)=r_{LL}(\varepsilon)r^*_{LL}(\varepsilon)$. Note that we have also the relation: $\mathcal{R}_{LL}(\varepsilon)=1-\mathcal{T}^{\text{eff},L}_{LR}(\varepsilon)$.

The terms appearing in Eq.~(\ref{MLL}) correspond precisely to the terms appearing in $|t_1+t_2+t_3|^2$, with $t_1=t_{LL}(\varepsilon)t_{LL}^*(\varepsilon-h\nu)$, $t_2=t_{LL}(\varepsilon)r_{LL}^*(\varepsilon-h\nu)$ and $t_3=r_{LL}(\varepsilon)t_{LL}^*(\varepsilon-h\nu)$, so that we finally obtain
\begin{eqnarray}
M_{LL}^{LL}(\varepsilon,\nu)&=&|t_1+t_2+t_3|^2~.
\end{eqnarray}


\section{C -- Non-interacting limit or elastic scattering limit}

In this Section, we show that when we neglect the interactions in the QD, or when we have only elastic scattering, the Table~I of Ref.~[\onlinecite{Zamoum2016}] can be derived from the Table~I given in the main text of this Letter. Indeed, in that case there are specific relations between the transmission and reflection coefficients which can be obtained from the optical theorem as summarized below.

\subsection{1 -- Optical theorem}

We define the $S$-matrix of a QD connected to a $L$-reservoir and a $R$-reservoir as ${\bf S}=\mathbb{1}+i{\bf T}$, where the $T$-matrix is given by
\begin{eqnarray}\label{tmatrix}
{\bf T}=\left(
\begin{array}{cc}
\tau_{LL}(\varepsilon)&\tau_{LR}(\varepsilon)\\
\tau_{RL}(\varepsilon)&\tau_{RR}(\varepsilon)
\end{array}
\right)~,
\end{eqnarray}
in the $\{L,R\}$ basis, where $\tau_{\alpha\beta}(\varepsilon)=it_{\alpha\beta}(\varepsilon)$ with $t_{\alpha\beta}(\varepsilon)$ the transmission amplitude from the $\alpha$-reservoir to the $\beta$-reservoir. For a QD, we have $\tau_{\alpha\beta}(\varepsilon)=-\sqrt{\Gamma_\alpha\Gamma_\beta}G_\sigma^r(\varepsilon)$, and thus $\tau_{RL}(\varepsilon)=\tau_{LR}(\varepsilon)$. At low temperature, when only elastic scattering of electrons are considered, ${\bf S}$ is a unitary matrix, ${\bf S}{\bf S}^+=\mathbb{1}$. Consequently the $T$-matrix must fulfill the optical theorem: ${\bf T}{\bf T}^+=i[{\bf T}^+-{\bf T}]$, leading to the following relations
\begin{eqnarray}\label{eq1}
2\mathrm{Im}\{\tau_{LL}(\varepsilon)\}&=&\mathcal{T}_{LL}(\varepsilon)+\mathcal{T}_{LR}(\varepsilon)~,\\\label{eq2}
2\mathrm{Im}\{\tau_{RR}(\varepsilon)\}&=&\mathcal{T}_{RR}(\varepsilon)+\mathcal{T}_{LR}(\varepsilon)~,\\
2\mathrm{Im}\{\tau_{LR}(\varepsilon)\}&=&\tau_{LR}^*(\varepsilon)\tau_{LL}(\varepsilon)+\tau_{LR}(\varepsilon)\tau_{RR}^*(\varepsilon)\nonumber\\\label{eq3}
&=&\tau_{LR}(\varepsilon)\tau_{LL}^*(\varepsilon)+\tau_{LR}^*(\varepsilon)\tau_{RR}(\varepsilon)~,
\end{eqnarray}
where we have defined the transmission coefficient as $\mathcal{T}_{\alpha\beta}(\varepsilon)=|\tau_{\alpha\beta}(\varepsilon)|^2=|t_{\alpha\beta}(\varepsilon)|^2$. Eqs.~(\ref{eq1}) to (\ref{eq3}) can equivalently written in terms of transmission amplitudes as
\begin{eqnarray}
2\mathrm{Re}\{t_{LL}(\varepsilon)\}&=&\mathcal{T}_{LL}(\varepsilon)+\mathcal{T}_{LR}(\varepsilon)~,\\
2\mathrm{Re}\{t_{RR}(\varepsilon)\}&=&\mathcal{T}_{RR}(\varepsilon)+\mathcal{T}_{LR}(\varepsilon)~,\\
2\mathrm{Re}\{t_{LR}(\varepsilon)\}&=&t_{LR}^*(\varepsilon)t_{LL}(\varepsilon)+t_{LR}(\varepsilon)t_{RR}^*(\varepsilon)\nonumber\\
&=&t_{LR}(\varepsilon)t_{LL}^*(\varepsilon)+t_{LR}^*(\varepsilon)t_{RR}(\varepsilon)~.
\end{eqnarray}
Note that these relations are automatically verified if the relation $G_\sigma^r(\varepsilon)-G_\sigma^a(\varepsilon)=-iG_\sigma^r(\varepsilon)(\Gamma_L+\Gamma_R)G_\sigma^a(\varepsilon)$ holds.

It is easy to show that provided the optical theorem holds, the reflection coefficient $\mathcal{R}_{\alpha\alpha}(\varepsilon)=|r_{\alpha\alpha}(\varepsilon)|^2$ with $r_{\alpha\alpha}(\varepsilon)=1+i\tau_{\alpha\alpha}(\varepsilon)=1-t_{\alpha\alpha}(\varepsilon)$, reads as
\begin{eqnarray}
\mathcal{R}_{LL}(\varepsilon)&=&\mathcal{R}_{RR}(\varepsilon)=1-\mathcal{T}_{LR}(\varepsilon)~.
\end{eqnarray}

\subsection{2 -- Matrix elements appearing in the noise}

When the optical theorem holds, which is the case in the non-interacting limit or in the elastic scattering processes limit, the matrix elements involved in Table I of this Letter can be rewritten as shown below in Table~\ref{table2}. Indeed, starting from the definition of the effective transmission coefficients, we have
\begin{eqnarray}
 \mathcal{T}_{LR}^{\text{eff},\alpha}(\varepsilon)&=&2\text{Re}\{t_{\alpha\alpha}(\varepsilon)\}-\mathcal{T}_{\alpha\alpha}(\varepsilon)=2\text{Im}\{\tau_{\alpha\alpha}(\varepsilon)\}-\mathcal{T}_{\alpha\alpha}(\varepsilon)=\mathcal{T}_{LR}(\varepsilon)~,
\end{eqnarray}
thanks to Eqs.~(\ref{eq1}) and (\ref{eq2}). Note that for symmetric couplings, i.e. $\Gamma_L=\Gamma_R$, the transmission amplitude does not depend on the reservoir index any longer, and is simply denoted $t(\varepsilon)$, which allows one to derive the matrix elements given in Table~I of Ref.~[\onlinecite{Zamoum2016}].

\begin{table}[h!]
\begin{center}
\begin{tabular}{|c||c|c|c|c|}
\hline
$M_{\alpha\beta}^{\gamma\delta}(\varepsilon,\nu)$& $\gamma=\delta=L$& $\gamma=\delta=R$&$\gamma=L$, $\delta=R$&$\gamma=R$, $\delta=L$\\ \hline\hline
$\alpha=L$& $\mathcal{T}_{LR}(\varepsilon)\mathcal{T}_{LR}(\varepsilon-h\nu)$& $\mathcal{T}_{LR}(\varepsilon)\mathcal{T}_{LR}(\varepsilon-h\nu)$ & $[1-\mathcal{T}_{LR}(\varepsilon)]\mathcal{T}_{LR}(\varepsilon-h\nu)$ & $\mathcal{T}_{LR}(\varepsilon)[1-\mathcal{T}_{LR}(\varepsilon-h\nu)]$\\
$\beta=L$&$+|t_{LL}(\varepsilon)-t_{LL}(\varepsilon-h\nu)|^2$&&&\\
 \hline
$\alpha=R$& $\mathcal{T}_{LR}(\varepsilon)\mathcal{T}_{LR}(\varepsilon-h\nu)$ &$\mathcal{T}_{LR}(\varepsilon)\mathcal{T}_{LR}(\varepsilon-h\nu)$ & $\mathcal{T}_{LR}(\varepsilon)[1-\mathcal{T}_{LR}(\varepsilon-h\nu)]$ & $[1-\mathcal{T}_{LR}(\varepsilon)]\mathcal{T}_{LR}(\varepsilon-h\nu)$ \\
$\beta=R$&&$+|t_{RR}(\varepsilon)-t_{RR}(\varepsilon-h\nu)|^2$&&\\
 \hline
$\alpha=L$& $t_{LR}(\varepsilon)t^*_{LR}(\varepsilon - h\nu)$ & $t^*_{LR}(\varepsilon)t_{LR}(\varepsilon - h\nu) $  & $t_{LR}(\varepsilon)t_{LR}(\varepsilon-h\nu)$ &$t_{LR}^*(\varepsilon)t_{LR}^*(\varepsilon-h\nu)$\\
$\beta=R$&$\times[r^*_{LL}(\varepsilon)r_{LL}(\varepsilon-h\nu)-1]$&$\times[r_{RR}(\varepsilon)r^*_{RR}(\varepsilon-h\nu)-1]$&$\times r_{LL}^*(\varepsilon)r_{RR}^*(\varepsilon-h\nu)$& $\times r_{RR}(\varepsilon)r_{LL}(\varepsilon-h\nu)$\\
 \hline
$\alpha=R$& $t^*_{LR}(\varepsilon)t_{LR}(\varepsilon - h\nu)$&$t_{LR}(\varepsilon)t^*_{LR}(\varepsilon - h\nu)$  &$t_{LR}^*(\varepsilon)t_{LR}^*(\varepsilon-h\nu)$&
$t_{LR}(\varepsilon)t_{LR}(\varepsilon-h\nu)$ \\
$\beta=L$&$\times[r_{LL}(\varepsilon)r^*_{LL}(\varepsilon-h\nu)-1]$&$\times[r^*_{RR}(\varepsilon)r_{RR}(\varepsilon-h\nu)-1]$&$\times r_{LL}(\varepsilon)r_{RR}(\varepsilon-h\nu)$& $\times r^*_{RR}(\varepsilon)r^*_{LL}(\varepsilon-h\nu)$\\
 \hline
\end{tabular}
\caption{Expressions of the matrix elements $M_{\alpha\beta}^{\gamma\delta}(\varepsilon, \nu)$ for a non-interacting QD, or in the presence of elastic scattering processes only.}
\label{table2}
\end{center}
\end{table}


\section{D -- Numerical calculation of $G_\sigma^r(\varepsilon)$ in the presence of Coulomb interactions}

When Coulomb interactions $U$ are present in the QD, it is necessary to take the spin degree of freedom into account. Indeed the Hamiltonian describing our system is the single-site Anderson Hamiltonian including on-site Coulomb interaction and reads as
\begin{eqnarray}
H=\sum_{k,\alpha \in (L,R) ,\sigma}{\varepsilon_{k\alpha}c_{k\alpha\sigma}^{\dag}c_{k\alpha\sigma}}+\sum_{\sigma}{\varepsilon_{0}d_{\sigma}^{\dag}d_{\sigma}}
+Un_{\uparrow}n_{\downarrow}+\sum_{k,\alpha \in (L,R) ,\sigma}{(V_{\alpha}c_{k\alpha\sigma}^{\dag}d_{\sigma}+h.c.)}~,
\end{eqnarray} 
where $d_\sigma^\dag$ ($c_{k\alpha\sigma}^\dag$) and $d_\sigma$ ($c_{k\alpha\sigma}$) are the creation and annihilation operators of an electron in the QD ($\alpha$-reservoir) respectively, and $n_\sigma=d^\dag_\sigma d_\sigma$.
Following Refs.~[\onlinecite{Roermund2010,Lavagna2015,Lavagna2017}], we numerically calculate the retarded Green function using the expression
\begin{eqnarray}
\label{fgreenfunction1}
G_{\sigma}^{r}(\varepsilon)&=&\frac{1-\langle n_{\bar{\sigma}} \rangle}{\varepsilon-\varepsilon_0-\Sigma_{\sigma}^0(\varepsilon)-\Pi_{\sigma}^{(1)}(\varepsilon)}
+\frac{\langle n_{\bar{\sigma}} \rangle} {\varepsilon-\varepsilon_0-U-\Sigma_{\sigma}^0(\varepsilon)-\Pi_{\sigma}^{(2)}(\varepsilon)},
\end{eqnarray}
with $\Sigma_{\sigma}^0(\varepsilon)=-i \Gamma_\sigma (\varepsilon)$ and $\Gamma_\sigma(\varepsilon)=\sum_{\alpha=L,R} \Gamma_{\alpha \sigma }(\varepsilon)$. In the FWBL, $\Sigma_{\sigma}^0(\varepsilon)$ is 
independent of $\varepsilon$ and takes the value $-i \Gamma_\sigma$. $\Pi_{\sigma}^{(1)}(\varepsilon)$ and $\Pi_{\sigma}^{(2)}(\varepsilon)$ are defined as
\begin{eqnarray}
\label{Pi1}
&\Pi_{\sigma} ^{(1)}(\varepsilon)&=-U \frac{\Sigma_{\sigma}^{(1)}(\varepsilon)-(\varepsilon-{\varepsilon}_0)\Sigma_{\sigma }^{(4)}(\varepsilon)}{\varepsilon-\varepsilon_0-U
-\Sigma_{\sigma}^{(3)}(\varepsilon)+U\Sigma_{\sigma}^{(4)}(\varepsilon)}~,\\
\label{Pi2}
&\Pi_{\sigma}^{(2)}(\varepsilon)&=U \frac{\Sigma_{\sigma}^{(2)}(\varepsilon)+(\varepsilon-{\varepsilon}_0-U)\Sigma_{\sigma}^{(4)}(\varepsilon)}{\varepsilon-\varepsilon_0
-\Sigma_{\sigma}^{(3)}(\varepsilon)+U\Sigma_{\sigma}^{(4)}(\varepsilon)}~,
\end{eqnarray}
where for $i\in[1,4]$,
\begin{eqnarray}
\label{Sigmai}
\Sigma_{\sigma}^{(i)}(\varepsilon)&=& \sum_{k,\alpha} { |V_{\alpha }|^2 \Big\lbrack} \frac{\mathcal{A}_{k\alpha\sigma}^{(i)}}{\varepsilon+ \widetilde{\varepsilon}_{\bar{\sigma}}-\widetilde
{\varepsilon}_\sigma-{\varepsilon}_{k \alpha}+i\widetilde{\gamma}_\sigma}
+\frac{\mathcal{A}_{k\alpha\sigma}^{\prime (i)}}{\varepsilon+\widetilde{\varepsilon}_{k \alpha}-\widetilde{\varepsilon}_\sigma- \widetilde{\varepsilon}_{\bar{\sigma}}-U+i\widetilde{\gamma}_D} { \Big\rbrack}~,
\end{eqnarray}
with $\widetilde {\varepsilon}_\sigma=\varepsilon_0+\mathrm{Re}\{\Sigma_{\sigma}^{(1)}(
\widetilde {\varepsilon}_\sigma)\}$, $\mathcal{A}_{k\alpha \sigma}^{(1)}= \sum_{k'} \langle c_{k'\alpha{\bar{\sigma}}}^\dagger  c_{k\alpha {\bar{\sigma}}} \rangle$, $\mathcal{A}_{k\alpha \sigma}^{(2)}=  1- \sum_{k'\alpha } \langle c_
{k'\alpha{\bar{\sigma}}}^\dagger  c_{k\alpha{\bar{\sigma}}} \rangle $, $\mathcal{A}_{k\alpha \sigma}^{(3)}=1$, and $\mathcal{A}_{k\alpha \sigma}^{(4)}=  \langle d_{\bar{\sigma}}^\dagger  c_{k\alpha{\bar
{\sigma}}} \rangle / V_{\alpha}$. $\mathcal{A}_{k\alpha \sigma}^{\prime (i)}=(\mathcal{A}_{k\alpha \sigma}^{(i)})^{*}$ for $i\in[1,3]$, and $\mathcal{A}_{k\alpha \sigma}^{\prime (4)}=-(\mathcal{A}_{k\alpha \sigma}^{(4)})
^{*}$. $\widetilde{\gamma}_\sigma$ and $\widetilde{\gamma}_D$ are calculated by using the Fermi golden rule up to the fourth order with $V_\alpha$ \cite{Lavagna2015,Lavagna2017}. The  numerical calculations are performed self-consistently.


\section{E -- Differential conductance in a Kondo QD}

\begin{figure}[!h]
\centering
\includegraphics[width=3.8cm]{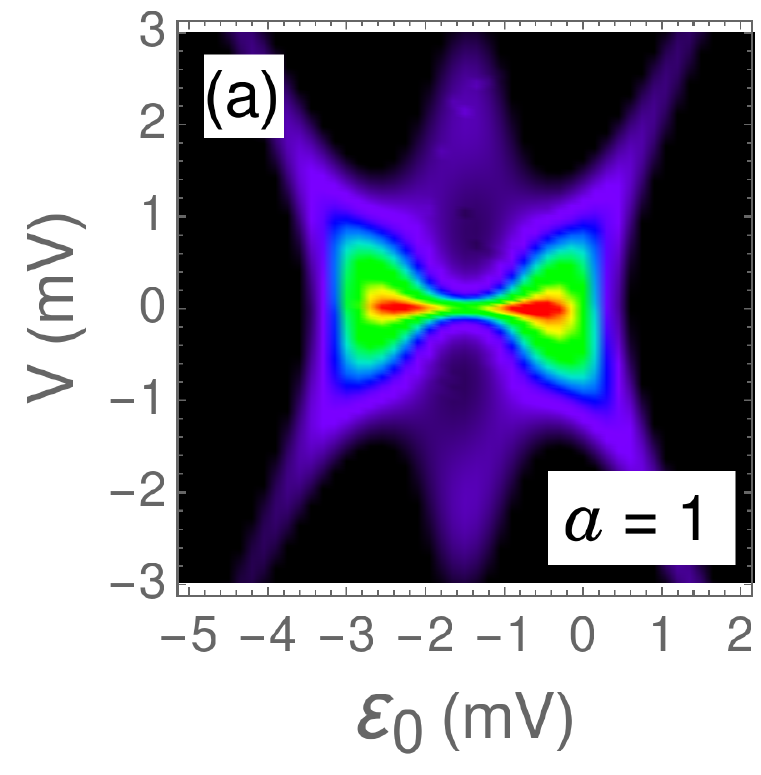}
\includegraphics[width=4.4cm]{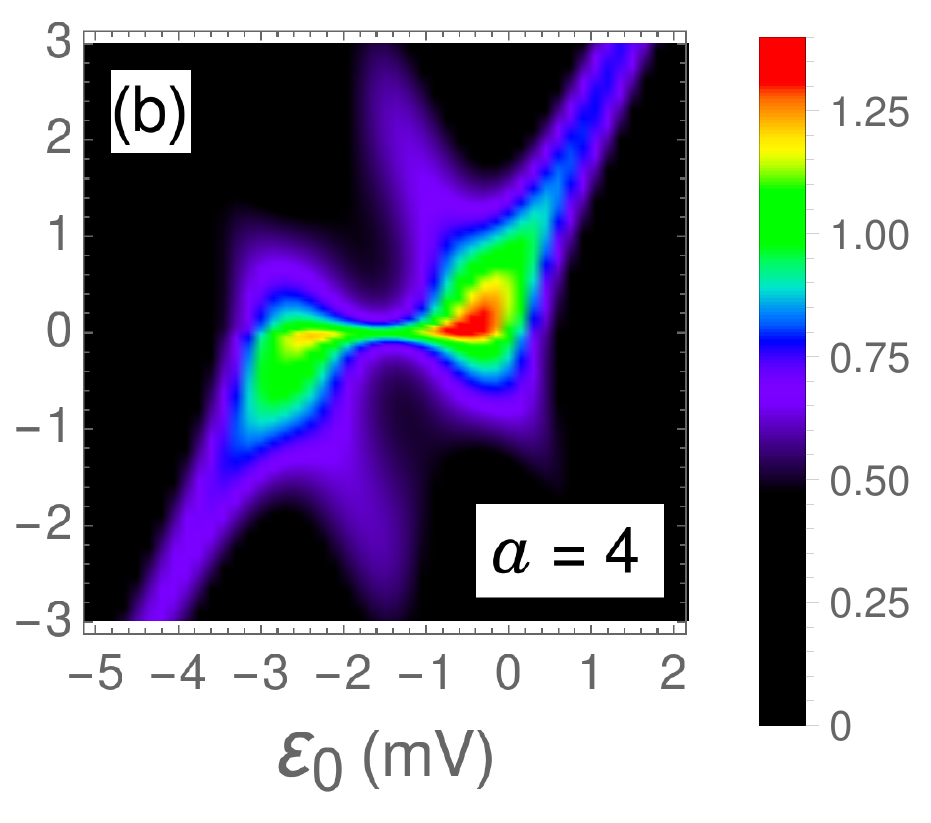}
\includegraphics[width=3.8cm]{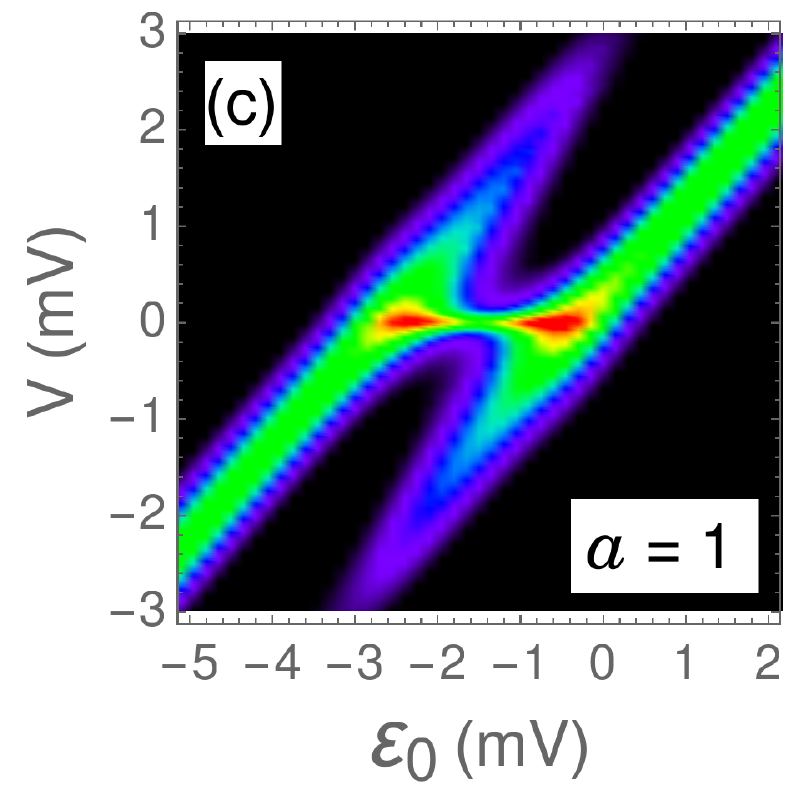}
\includegraphics[width=4.4cm]{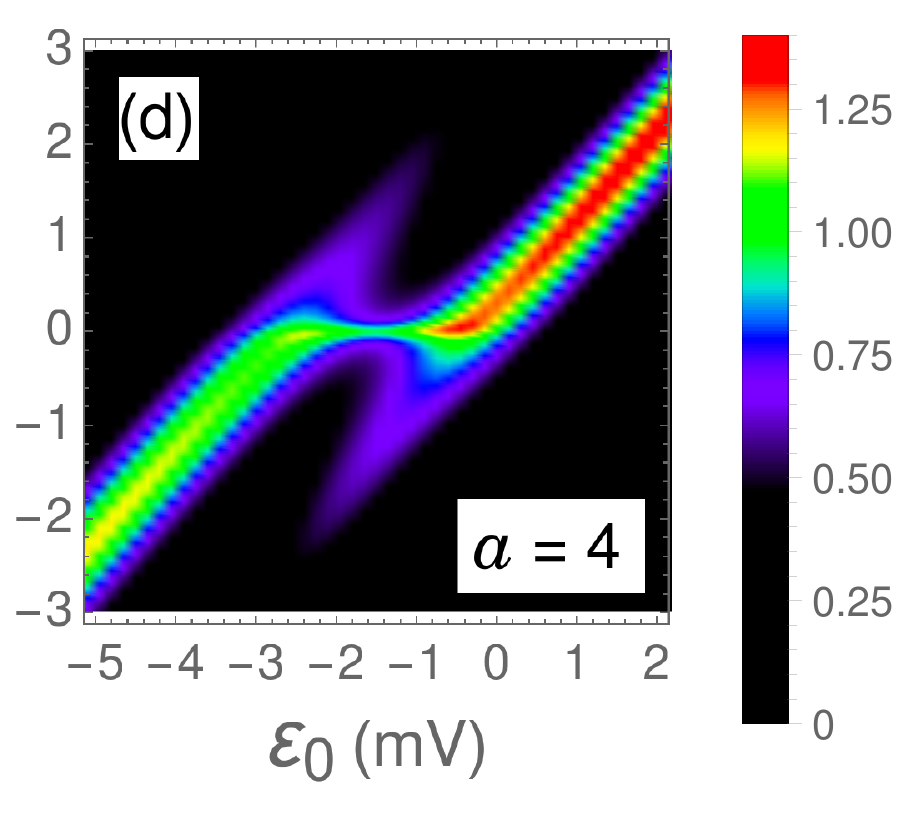}
\caption{Color-plot of the differential conductance $G$ of an interacting QD (in units of $e^2/h$) as a function of the level energy $\varepsilon_0$ and bias voltage $V$, for $T=80$ mK and $U=3$ meV. (a) and (c): symmetric couplings $\Gamma_{L,R}=0.5$ meV ($a=1$). (b) and (d): asymmetric couplings $\Gamma_L=0.8$ meV, $\Gamma_R=0.2$ meV ($a=4$). (a) and (b): symmetric bias voltage profile {$\mu_L=-\mu_R=-eV/2$}. (c) and (d): asymmetric bias voltage profile $\mu_L=0$ and $\mu_R=eV$.}
\label{figureS1}
\end{figure}

In Fig.~\ref{figureS1}, we report the color-plot of the differential conductance, $G=dI/dV$, for an interacting QD as a function of the level energy $\varepsilon_0$ and bias voltage $V$.  The results are shown for both symmetric and asymmetric couplings, and both symmetric and asymmetric bias voltage profiles. In the presence of Coulomb interactions ($U=3$ meV), we observe a Kondo ridge in the $n=1$ conductance valley ($n$ being the QD occupation) characteristic of the Kondo effect which manifests itself at low temperature  {(see plots in Fig.~\ref{figureS1} at $V=0$)}. Here $T=80$ mK is lower than the Kondo temperature, $T_K\approx 4.38$~K, estimated from the Haldane formula: $k_BT_K\approx \sqrt{U\Gamma/2}\exp(\pi\varepsilon_0(\varepsilon_0+U)/2U\Gamma)$, with $\Gamma=\Gamma_L+\Gamma_R$. Moreover, one observes a Coulomb blockade structure: a Coulomb diamond (shown in violet in the center of the Fig.~\ref{figureS1}(a)) when the bias voltage profile is symmetric, or two parallel branches ($eV=\varepsilon_0$ 
and $eV=\varepsilon_0+U$) when the bias voltage profile is asymmetric (see Fig.~\ref{figureS1}(c)). The introduction of an asymmetry in the couplings (weakening $\Gamma_R$ over $\Gamma_L$, keeping $\Gamma_R+\Gamma_L$ constant) induces the following two changes in the conductance: (i)~the value of the conductance decreases due to the reduced transmission through the QD, as it can be clearly seen by comparing the intensity along the Kondo ridge; (ii)~at a given value of $\varepsilon_0$, the relative height of the two broad peaks  {is} changed, e.g., in case of asymmetric bias voltage profile  {(Fig.~\ref{figureS1}(d))}, the two sub-branches, $eV=\varepsilon_0+U<0$ and $eV=\varepsilon_0>0$, becomes more prominent than the two other sub-branches $eV=\varepsilon_0+U>0$ and $eV=\varepsilon_0<0$.

Additional remark: the profile of the bias voltage through the QD and the asymmetry of the left and right couplings are closely linked. It is generally expected that the profile is symmetric at $a=1$, and asymmetric at $a\ne 1$. However only the consideration of the electrodynamics of the whole system can help in determining the link. Here, we rather consider all the possible cases in Fig.~\ref{figureS1}, but choose to focus on an asymmetric bias voltage profile to be able to compare the curves of Fig.~2 given in the main text of this Letter with experiments of Ref.~[\onlinecite{Delagrange2017}].

\end{document}